\documentclass[preprint,showpacs,superscriptaddress]{revtex4}
\usepackage{graphicx}

\usepackage{amsmath,amssymb,amsfonts}
\usepackage{floatflt}
\usepackage{wrapfig}
\usepackage{indentfirst}

\begin{document}

\title{The quantization of a charge qubit. The role of inductance and gate capacitance}
\author{Ya. S. Greenberg }
\affiliation{Novosibirsk State Technical University, 20 K. Marx
Ave., 630092 Novosibirsk, Russia}

\author{W. Krech}
\affiliation{Friedrich-Schiller University, Jena, Germany}

\date{\today}

\begin{abstract}

The Hamiltonian of a charge qubit, which consists of two Josephson
junctions is found within well known quantum mechanical procedure.
The inductance of the qubit is included from the very beginning.
It allows a selfconsistent derivation of the current operator in a
two state basis. It is shown that the current operator has nonzero
nondiagonal matrix elements both in the charge and the eigenstate
basis. It is also shown that the interaction of the qubit with its
own LC resonator has a noticeable influence on the qubit energies.
The influence of the junctions asymmetry and the gate capacitance
on the matrix elements of the current operator and on the qubit
energies are calculated. The results obtained in the paper are
important for the circuits where two or more charge qubits are
coupled with the aid of inductive coil.

\end{abstract}

\maketitle

\section{Introduction}
Josephson-junction charge qubits are known to be candidates for
scalable solid-state quantum computing circuits \cite{Makhlin},
\cite{Nak}, \cite{Makhlin1}. Here we consider a superconducting
charge qubit which consists of two Josephson junctions embedded in
a loop with very small inductance~$L$, typically in the pH range.
This insures effective decoupling from the environment. However,
in the practical implementation of qubit circuitry it is important
to have the loop inductance as much as possible consistent with a
proper operation of a qubit. A relative large loop inductance
facilitates a qubit control biasing schemes and the formation,
control and readout of two-qubit quantum gates. These
considerations stimulated some investigations of the role the loop
inductance plays in the dynamic properties of charge qubits
\cite{You}, \cite{You1}, \cite{Krech}, where for the small loop
inductance the corrections to the energy levels due to finite
inductance of the loop have been found. The corrections have been
obtained by perturbation expansion of the energy over small
parameter $\beta=L/L_J$, where $L_J$ is the Josephson junction
inductance.

For complex Josephson circuit the construction of quantum
Hamiltonian which accounts for finite inductances of
superconducting loops can be made with the aid of the graph theory
\cite{Dev}. This approach has been developed in \cite{Bur1} for
systematic derivation of the Hamiltonian of superconducting
circuits and has been applied for the calculations of the effects
of the finite loop inductance both for flux \cite{Bur2} and charge
\cite{Bur3} qubits.

In principle, the account for a finite loop inductance (even if it
is small) requires for the magnetic energy to be included in
quantum mechanical Hamiltonian of a qubit from the very beginning.
It allows one to obtain the effects of the interaction between
two-level qubit and its own LC circuit. In addition it allows a
correct definition of the current operator in terms of its matrix
elements in a two level basis.

In this paper we investigate the effect of finite loop inductance
and gate capacitance for a asymmetric charge qubit, which consists
of two Josephson junctions embedded in a superconducting loop.

The construction of exact Lagrangian and Hamiltonian for the
charge qubit is given in Section II and Section III, respectively.
The approximation for exact Hamiltonian for small $L$ is made in
Section IV. It is shown that Hamiltonian is decomposed in three
parts: qubit part, LC-oscillator part and the qubit-LC oscillator
interaction part. In this approximation the energy levels of the
charge qubit are explicitly dependent on the gate capacitance and
critical current asymmetry and , in addition, are shifted due to
vacuum fluctuations of LC oscillator. The current operator both in
charge and in eigenstate basis is obtained in Section V. It is
shown that the asymmetry of critical currents of the Josephson
junctions results in additional terms in the operator of critical
current. The corrections to the qubit energies due to its
interaction with LC circuit and their dependence on critical
current asymmetry and on gate capacitance are calculated in
Section VI.

\section{Lagrangian for the charge qubit}
\begin{figure}[h]
\centerline{\includegraphics[width=8cm, angle=-90]{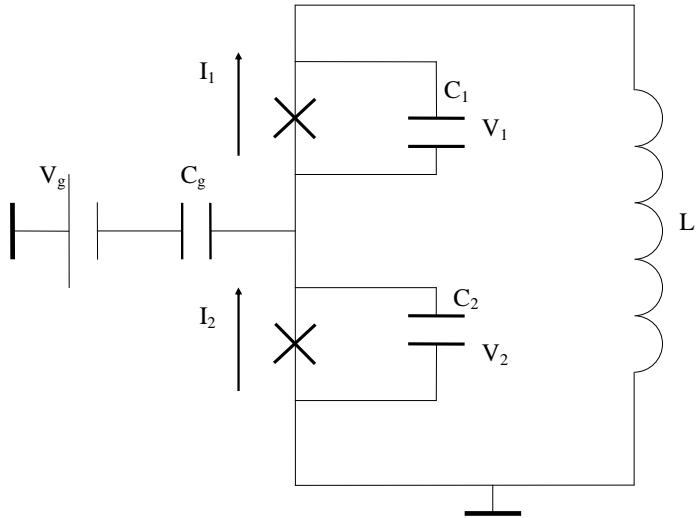}}
\caption{Charge qubit with inductance coil.} \label{fig1}
\end{figure}

We consider here a charge qubit in the arrangement, which has been
first proposed in \cite{Makhlin} (see Fig.\ref{fig1}).
The qubit consists of two Josephson junctions in a loop with very
small inductance~$L$, typically in the pH range. This insures
effective decoupling from the environment. As a general case we
assume that two junctions have different critical currents
$I_\mathrm{c1}$, $I_\mathrm{c2}$ and capacitance~$C_1$, $C_2$. The
Josephson energy $E_\mathrm{J}=I_\mathrm{c}\Phi_0/2\pi$, where
$\Phi_0=h/2e$ is the flux quantum, is assumed to be much less than
the Coulomb energy $E_C=(2e)^2\!/2C$, so that the charge at the
gate is well defined.

The Lagrangian of this qubit is the difference between the charge
energy in the junction capacitors and the Josephson plus magnetic
energy:
\begin{equation}\label{Lag1}
{\rm{L}} = U - \frac{{\Phi ^2 }}{{2L}} + E_{J1} \cos \varphi _1  +
E_{J2} \cos \varphi _2
\end{equation}
where $U$ is the electric energy of JJ's and gate capacities
\begin{equation}\label{U}
U = \frac{{C_1 V_1^2 }}{2} + \frac{{C_2 V_2^2 }}{2} + \frac{{C_g
V^2 }}{2},
\end{equation}
$\Phi  = \int {V_L dt} $, where $V_L$ is the voltage drop across
the inductance.

In virtue  of Josephson relations
\begin{equation}\label{JosRel}
V_i  = \frac{\hbar }{{2e}}\dot \varphi _i \;,\quad i = 1,2;\quad V
= V_g  + \frac{\hbar }{{2e}}\dot \varphi _2 \,
\end{equation}
the voltage drop across the inductance is:
\begin{equation}\label{VL}
V_L  = V_1  + V_2  - \frac{{d\Phi _X }}{{dt}} =
\frac{d}{{dt}}\left[ {\frac{\hbar }{{2e}}\left( {\varphi _1  +
\varphi _2 } \right) - \Phi _X } \right]
\end{equation}
where $\Phi_X$ is the external flux.

In terms of the phases $\varphi_1$, $\varphi_2$, $\varphi_3$
Lagrangian (\ref{Lag1}) takes the form:

\begin{eqnarray}\label{Lag2}
{\rm{L}} = \frac{{\hbar ^2 }}{{2(2e)^2 }}\left( {C_1 \dot \varphi
_1^2  + C_2 \dot \varphi _2^2 } \right) + \frac{{C_g }}{2}\left(
{V_g  + \frac{\hbar }{{2e}}\dot \varphi _2 } \right)^2 \\\nonumber
- \frac{{\hbar ^2 }}{{(2e)^2 }}\frac{{\left( {\varphi _1  +
\varphi _2  - \varphi _X } \right)^2 }}{{2L}} + E_{J1} \cos
\varphi _1 + E_{J2} \cos \varphi _1
\end{eqnarray}
where $\varphi_X=2\pi\Phi_X/\Phi_0$.

Next  we make the known (Likharev and Averin) redefinition of the
Josephson phases: $\varphi_1+\varphi_2=\varphi$;
$\varphi_1-\varphi_2=2\delta$. Lagrangian (\ref{Lag2}) takes the
form:
\begin{eqnarray}\label{Lag3}
{\rm{L}} = \frac{{\hbar ^2 \left( {C_1  + C_2 } \right)}}{{2(2e)^2
}}\left( {\frac{{\dot \varphi ^2 }}{4} + \dot \delta ^2 } \right)
+ \frac{{\hbar ^2 \left( {C_1  - C_2 } \right)}}{{2(2e)^2 }}\dot
\delta \dot \varphi  + \frac{{C_g }}{2}\left( {V_g  + \frac{\hbar
}{{4e}}\dot \varphi  - \frac{\hbar }{{2e}}\dot \delta }
\right)^2\\\nonumber - \frac{{\hbar ^2 }}{{(2e)^2 }}\frac{{\left(
{\varphi  - \varphi _X } \right)^2 }}{{2L}} + \left( {E_{J1}  +
E_{J2} } \right)\cos \frac{\varphi }{2}\cos \delta  + \left(
{E_{J2} - E_{J1} } \right)\sin \frac{\varphi }{2}\sin \delta
\end{eqnarray}
\section{Construction of Hamiltonian}
 Conjugate variables are defined in a standard way:
\begin{equation}\label{nphi}
n_\varphi   = \frac{1}{\hbar }\frac{{\partial L}}{{\partial \dot
\varphi }} = \frac{{\hbar \left( {C_1  + C_2  + C_g }
\right)}}{{4(2e)^2 }}\dot \varphi  + \frac{{\hbar \left( {C_1 -
C_2  - C_g } \right)}}{{2(2e)^2 }}\dot \delta  + \frac{{C_g V_g
}}{{4e}}
\end{equation}
\begin{equation}\label{ndelta}
n_\delta   = \frac{1}{\hbar }\frac{{\partial L}}{{\partial \dot
\delta }} = \frac{{\hbar \left( {C_1  + C_2  + C_g }
\right)}}{{(2e)^2 }}\dot \delta  + \frac{{\hbar \left( {C_1  - C_2
- C_g } \right)}}{{2(2e)^2 }}\dot \varphi  - \frac{{C_g V_g
}}{{2e}}
\end{equation}
From these equations we express phases in terms of conjugate
variables:
\begin{equation}\label{phidot}
\dot \varphi  = \frac{{E_C \alpha }}{\hbar }\left[ {2\left(
{n_\varphi   - \frac{{n_g }}{2}} \right) + \gamma \left( {n_\delta
+ n_g } \right)} \right]
\end{equation}
\begin{equation}\label{deltadot}
\dot \delta  = \frac{{E_C \alpha }}{\hbar }\left[ {\gamma \left(
{n_\varphi   - \frac{{n_g }}{2}} \right) + \frac{1}{2}\left(
{n_\delta   + n_g } \right)} \right]
\end{equation}
where $E_C=(2e)^2/2C_\Sigma$, $\alpha = C_\Sigma ^2/C_1(C_2 +
C_g)$, $\gamma = (C_g  + C_2  - C_1)/C_\Sigma$,
$C_\Sigma=C_1+C_2+C_g$, $n_g=C_gV_g/2e$.

Now we construct Hamiltonian:
\begin{equation}\label{Ham1}
{\rm{H = }}\hbar n_\varphi  \dot \varphi  + \hbar n_\delta  \dot
\delta  - {\rm{L}}
\end{equation}
Eliminating time derivatives of the phases from (\ref{Ham1}) with
the aid of (\ref{deltadot}), (\ref{phidot}), we obtain  the final
expression for Hamiltonian of the asymmetric charge qubit:
\begin{eqnarray}\label{Ham2}
{\rm{H}} = E_C \alpha \left( {n_\varphi   - \frac{{n_g }}{2}}
\right)^2  + \frac{{E_C \alpha }}{4}\left( {n_\delta   + n_g }
\right)^2 + E_C \alpha \gamma \left( {n_\varphi   - \frac{{n_g
}}{2}} \right)\left( {n_\delta  + n_g } \right) -
\\\nonumber
 {\rm{}} - 2E_J \cos \frac{\varphi }{2}\cos \delta
  - E_J \xi \sin \frac{\varphi }{2}\sin \delta
  + E_J\frac{{\left( {\varphi - \varphi _X } \right)^2 }}{{2\beta}}
  - \frac{{(2e)^2 }}{{2C_g }}n_g^2
\end{eqnarray}
where $E_J=\Phi_0 I_C/2\pi$, $I_C=(I_{C1}+I_{C2})/2$,
$\xi=(I_{C2}-I_{C1})/I_C$, $\beta=2\pi LI_C/\Phi_0$.

The first two equations of motion
\[
\dot \delta  = \frac{1}{\hbar }\frac{{\partial H}}{{\partial
n_\delta  }};\quad \dot \varphi  = \frac{1}{\hbar }\frac{{\partial
H}}{{\partial n_\varphi  }}\] are given by Eqs. (\ref{phidot}) and
(\ref{deltadot}). Two other equations are as follows:
\begin{equation}\label{ndeltadot}
\dot n_\delta   =  - \frac{1}{\hbar }\frac{{\partial H}}{{\partial
\delta }} =  - \frac{{2E_J }}{\hbar }\cos \frac{\varphi }{2}\sin
\delta  + \frac{{E_J }}{\hbar }\xi \sin \frac{\varphi }{2}\cos
\delta
\end{equation}

\begin{equation}\label{nphidot}
\dot n_\varphi   =  - \frac{1}{\hbar }\frac{{\partial
H}}{{\partial \varphi }}= - \frac{{E_J }}{\hbar }\sin
\frac{\varphi }{2}\cos \delta  + \frac{{E_J }}{2\hbar }\xi \cos
\frac{\varphi }{2}\sin \delta  - \frac{E_J }{\hbar}\frac{{\varphi
- \varphi _X }}{\beta}
\end{equation}
Below we consider Hamiltonian (\ref{Ham2}) as quantum mechanical
with commutator relations imposed on its variables
\begin{equation}\label{Comm1}
\left[ {\varphi ,n_\varphi  } \right] = i;\quad \left[ {\delta
,n_\delta  } \right] = i
\end{equation}
where $n_\varphi=-i\partial/\partial\varphi$,
$n_\delta=-i\partial/\partial\delta$.

\section{Approximation to quantum mechanical Hamiltonian}
Obviously, Hamiltonian (\ref{Ham2}) is 2D nonlinear oscillator. We
assume L is small, so that its frequency $\left( {LC_\Sigma  }
\right)^{ - 1/2}  >  > E_J /\hbar $ . Therefore we can consider
$\varphi$ as fast variable with fast oscillations near the point
$\varphi_C$, the minimum of potential $U(\varphi,\delta )$ (see
(\ref{Ham2})):
\begin{eqnarray}\label{Pot}
 U(\varphi,\delta)= - 2E_J \cos \frac{\varphi }{2}\cos \delta
  - E_J \xi \sin \frac{\varphi }{2}\sin \delta
  + E_J\frac{{\left( {\varphi - \varphi _X } \right)^2 }}{{2\beta}}
\end{eqnarray}

We single out of this potential the fast variable $\varphi$, which
describes the interaction of the qubit with its own $LC$ circuit.

The point of minimum $\varphi_C$ of $U(\varphi, \delta)$
(\ref{Pot}) with respect to $\varphi$ is defined from $\partial
U/\partial\phi=0$:
\begin{equation}\label{phiC}
\varphi _C  = \varphi _X  - \beta \sin \frac{{\varphi _C }}{2}\cos
\delta  + \frac{{\beta \xi }}{2}\cos \frac{{\varphi _C }}{2}\sin
\delta
\end{equation}
In what follows we consider $\delta$ as slow variable and expand
$U(\varphi,\delta)$ near the point of minimum to the third order
in $\varphi$ ($\varphi=\varphi_C+\widehat{\varphi}$). In the
vicinity of $\varphi_C$ the potential $U(\varphi,\delta)$ can be
written as:
\begin{eqnarray}\label{Pot1}
U(\varphi ,\delta ) = U(\varphi _C ,\delta ) + \frac{{E_J
}}{{2\beta }}\left( {1 + \frac{\beta }{2}\cos \frac{{\varphi _C
}}{2}\cos \delta  + \frac{{\beta \xi }}{4}\sin \frac{{\varphi _C
}}{2}\sin \delta }
\right)\mathord{\buildrel{\lower3pt\hbox{$\scriptscriptstyle\frown$}}
\over \varphi } ^2\\\nonumber - \frac{{E_J
}}{{24}}\mathord{\buildrel{\lower3pt\hbox{$\scriptscriptstyle\frown$}}
\over \varphi } ^3 \left( {\sin \frac{{\varphi _C }}{2}\cos \delta
- \frac{\xi }{2}\cos \frac{{\varphi _C }}{2}\sin \delta } \right)
\end{eqnarray}
where $\widehat{\varphi}$ is the operator conjugate to
$n_\varphi$.

With the aid of (\ref{phiC}) we write $U(\varphi_C,\delta)$ to the
first order in $\beta$:
\begin{eqnarray}\label{Pot2}
 U(\varphi _C ,\delta ) \equiv U(\varphi _X ,\delta )=
 - 2E_J \cos \frac{{\varphi _X }}{2}\cos \delta
  - E_J \xi \sin \frac{{\varphi _X }}{2}\sin \delta  \\\nonumber
  - \frac{{E_J \beta }}{2}\left( {\sin ^2 \frac{{\varphi _X }}{2}\cos ^2 \delta  -
  \frac{\xi }{4}\sin \varphi _X \sin 2\delta  +
  \frac{{\xi ^2 }}{4}\cos ^2 \frac{{\varphi _X }}{2}\sin ^2 \delta } \right)
 \end{eqnarray}
Therefore, we decompose Hamiltonian (\ref{Ham2}) into oscillator,
qubit and interaction parts: $H=H_{osc}+H_{qb}+H_{int}$, where
\begin{equation}\label{Hosc}
H_{osc}  = E_C \alpha n_\varphi ^2  + \frac{{E_J }}{{2\beta
}}\mathord{\buildrel{\lower3pt\hbox{$\scriptscriptstyle\frown$}}
\over \varphi } ^2  - E_C \alpha (1 - \gamma )n_g n_\varphi
 {\rm{         }}
\end{equation}
\begin{equation}\label{Hqb}
H_{qb}  = \frac{{E_C \alpha }}{4}\left( {n_\delta   + n_g }
\right)^2  - \frac{{E_C \alpha }}{2}\gamma n_g \left( {n_\delta +
n_g } \right){\rm{ + U(}}\varphi _{\rm{X}} {\rm{,}}\delta {\rm{)}}
\end{equation}
\begin{eqnarray}\label{Hint}
 H_{{\mathop{\rm int}} }  = E_C \alpha \gamma n_\varphi  n_\delta
   + \frac{{E_J }}{4}\mathord{\buildrel{\lower3pt\hbox{$\scriptscriptstyle\frown$}}
\over \varphi } ^2 \left( {\cos \frac{{\varphi _X }}{2}\cos \delta
 + \frac{\xi }{2}\sin \frac{{\varphi _X }}{2}\sin \delta } \right)
 \\\nonumber
  - \frac{{E_J }}{{24}}\mathord{\buildrel{\lower3pt\hbox{$\scriptscriptstyle\frown$}}
\over \varphi } ^3 \left( {\sin \frac{{\varphi _X }}{2}\cos \delta
  - \frac{\xi }{2}\cos \frac{{\varphi _X }}{2}\sin \delta } \right)
  \\\nonumber
  + \beta \frac{{E_J }}{8}\mathord{\buildrel{\lower3pt\hbox{$\scriptscriptstyle\frown$}}
\over \varphi } ^2 \left( {\sin ^2 \frac{{\varphi _X }}{2}\cos ^2 \delta
 - \frac{\xi }{8}\sin \varphi _X \sin 2\delta  + \frac{{\xi ^2 }}{8}\cos ^2
 \frac{{\varphi _X }}{2}\sin ^2 \delta } \right) \\\nonumber
  + \beta \frac{{E_J }}{{96}}\mathord{\buildrel{\lower3pt\hbox{$\scriptscriptstyle\frown$}}
\over \varphi } ^3 \left( {\sin \varphi _X \cos ^2 \delta  -
\frac{\xi }{2}\cos \varphi _X \sin 2\delta  - \frac{{\xi ^2
}}{4}\sin \varphi _X \sin ^2 \delta } \right)
 {\rm{         }}
\end{eqnarray}
In the above equations we disregard the constant term which is
proportional to $n_g^2$.

The first term in (\ref{Hint}) describes the interaction of the
phase variables of the qubit, $\varphi$ and $\delta$ via the gate,
the other terms are responsible for the interaction of the qubit
with its own LC circuit.
\subsection{Two-level approximation}
First we quantize (\ref{Hosc}) according to ($n_\varphi   =  -
i\frac{\partial }{{\partial \varphi }};\;[n_\varphi  ,\varphi ] =
- i$):
\begin{equation}\label{Quant}
\varphi  = \frac{1}{{\sqrt 2 }}\left( {\frac{{2\beta E_C \alpha
}}{{E_J }}} \right)^{1/4} \left( {a^ +   + a} \right);\;\quad
n_\varphi   = i\frac{1}{{\sqrt 2 }}\left( {\frac{{E_J }}{{2\beta
E_C \alpha }}} \right)^{1/4} \left( {a^ +   - a} \right)
\end{equation}
where $[a,a^+]=1$.

In addition, we use the two level approximation in the charge
basis: $n_\delta   = \frac{1}{2}\left( {1 + \tau _Z } \right)$;
 $\cos \delta  = \tau _X/2$; $\sin \delta  = \tau _Y/2$ with Pauli
 operators
\begin{eqnarray}
\tau _Z \left| 0 \right\rangle = - \left| 0 \right\rangle; \tau _Z
\left| 1 \right\rangle = \left| 1 \right\rangle ;\\\nonumber\tau
_X \left| 0 \right\rangle  = \left| 1 \right\rangle;\tau _X \left|
1 \right\rangle  = \left| 0 \right\rangle;\\\nonumber \tau _Y
\left| 0 \right\rangle  = -i\left| 1 \right\rangle; \tau _Y \left|
1 \right\rangle  = i\left| 0 \right\rangle.
\end{eqnarray}
In this approximation $\sin2\delta$=$\cos2\delta=0$, since these
operators couple charge states which differs by two Cooper pairs.
Therefore, $\cos^2\delta$=$\sin^2\delta$=$1/2$.

Now we write down Hamiltonian (\ref{Hosc},\ref{Hqb}, \ref{Hint})
within two level subspace in terms of Pauli operators
$\tau_X,\tau_Y,\tau_Z$ and oscillator operators $a^+,a$.

We obtain the following result:
\begin{equation}\label{HamTL}
    H=W_0+H_{osc}+H_{qb}+H_{int}
\end{equation}
where
\begin{equation}\label{W0}
    W_0=\frac{E_C\beta\xi^2}{32}\cos\varphi_X
\end{equation}
\begin{eqnarray}\label{HamTLosc}
H_{osc}  = E_0 \left( {a^ +  a + \frac{1}{2}} \right) +
 i\frac{{E_C \alpha }}{{\sqrt 2\eta }}\left[ {\frac{\gamma }{2} -
 (1 - \gamma )n_g } \right]\left( {a^ +   - a} \right)\\\nonumber
  + \frac{{E_J \beta }}{{64}}\eta ^2 \left( {\left( {1 +
  \frac{{\xi ^2 }}{8}} \right) - \left( {1 - \frac{{\xi ^2 }}{8}} \right)\cos \varphi _X }
  \right)\left( {a^ +   + a} \right)^2 \\\nonumber +
  \frac{{E_J \beta }}{{384\sqrt 2 }}\eta ^3 \left( {1 - \frac{{\xi ^2 }}{4}} \right)
  \sin \varphi _X \left( {a^ +   + a} \right)^3
\end{eqnarray}
\begin{equation}\label{HamTLqb}
H_{qb}  = \frac{{E_C \alpha }}{8}\left( {1 + 2(1 - \gamma )n_g }
\right)\tau _Z  - \tau _X E_J \cos \frac{{\varphi _X }}{2} - \tau
_Y E_J \frac{\xi }{2}\sin \frac{{\varphi _X }}{2}
\end{equation}
\begin{eqnarray}\label{HamTLint}
 H_{int} = \frac{i}{{2^{3/2} }}\frac{{E_C \alpha \gamma }}{\eta }\left(
 {a^ +   - a} \right)\tau _Z  + E_J \frac{{\eta ^2 }}{{16}}\left(
 {\tau _X \cos \frac{{\varphi _X }}{2} + \tau _Y \frac{\xi }{2}\sin \frac{{\varphi _X }}{2}}
  \right)\left( {a^ +   + a} \right)^2  \\\nonumber
  - E_J \frac{{\eta ^3 }}{{96\sqrt 2 }}\left( {\tau _X \sin \frac{{\varphi _X }}{2} -
  \tau _Y \frac{\xi }{2}\cos \frac{{\varphi _X }}{2}} \right)\left( {a^ +   + a} \right)^3
\end{eqnarray}
where
\[
E_0  = \left( {\frac{{2E_C E_J \alpha }}{\beta }} \right)^{1/2}
,\quad \eta  = \left( {\frac{{2\beta E_C \alpha }}{{E_J }}}
\right)^{1/4} \]
\subsection{The energy levels of the charge qubit}
Here we neglect the interaction of the qubit with its own LC
circuit. It is justified if $\beta$ is sufficiently small so that
the energy levels of the qubit oscillator are located much higher
than the ground level of the qubit. The approximation we make here
is to average Hamiltonian (\ref{HamTL}) over the vacuum state,
$a^+a=0$, of the qubit oscillator. The result is as follows:
\begin{equation}\label{H}
    H=W+\frac{1}{2}A\tau_X+\frac{1}{2}B\tau_Y+\frac{1}{2}C\tau_Z
\end{equation}
where
\begin{equation}\label{W}
W = \frac{\beta }{{32}}\left[ {E_C \xi ^2  - \frac{{E_J \eta ^2
}}{2}\left( {1 - \frac{{\xi ^2 }}{8}} \right)} \right]\cos \varphi
_X
\end{equation}
\begin{equation}\label{A}
A= - 2E_J \left( {1 - \frac{{\eta ^2 }}{{16}}} \right)\cos
\frac{{\varphi _X }}{2}
\end{equation}
\begin{equation}\label{B}
B=- E_J \xi\left( {1 - \frac{{\eta ^2 }}{{16}}} \right)\sin
\frac{{\varphi _X }}{2}
\end{equation}

\begin{equation}\label{C}
C= \frac{{E_C \alpha }}{4}\left[ {1 + 2(1 - \gamma )n_g } \right]
\end{equation}

Hamiltonian (\ref{H}) has the corrections on the order of
$\eta^2\approx\sqrt{\beta}$ which are due to the vacuum
fluctuations of the LC oscillator. Since in a charge qubit
$E_C>>E_J$ these corrections in principle might be not very small.

Hamiltonian (\ref{H}) can be made diagonal in the eigenbasis with
the aid of the matrix \cite{Krech}:
\begin{equation}\label{S1}
\widehat{S}= \left( {\begin{array}{*{20}c}
   { - e^{ - i\Psi } \sin \frac{\theta }{2}} & {\cos \frac{\theta }{2}}  \\
   {\cos \frac{\theta }{2}} & {e^{i\Psi } \sin \frac{\theta }{2}}  \\
\end{array}} \right)
\end{equation}
where $\sin\theta=\varepsilon/\Delta E$, $\cos\theta=C/\Delta E$,
$\sin\Psi=B/\varepsilon$, $\cos\Psi=A/\epsilon$;
$\varepsilon=\sqrt{A^2+B^2}$, $\Delta E=\sqrt{\varepsilon^2+C^2}$.

The qubit Hamiltonian in eigenstate basis, therefore, reads:
\begin{equation}\label{HamEig}
    \widehat{S}^{-1}H\widehat{S}=W-\frac{1}{2}\Delta E\sigma_Z
\end{equation}
where $W$ is given by (\ref{W}) and
\begin{equation}\label{DE}
\Delta E = \sqrt {4E_J^2 \left( {1 - \frac{{\eta ^2 }}{{16}}}
\right)^2 \left( {\cos ^2 \frac{{\varphi _X }}{2} + \frac{{\xi ^2
}}{4}\sin ^2 \frac{{\varphi _X }}{2}} \right) + \left( {\frac{{E_C
\alpha }}{4}} \right)^2 \left[ {1 + 2n_g (1 - \gamma )} \right]^2
}
\end{equation}
As is seen from (\ref{DE}) the energies of the charge qubit
account for its full asymmetry and are explicitly dependent on the
inductance and on the gate capacitance.
\section{Current operator}
From the first principles the average current in the loop is equal
to the first derivative of the eigenenergy relative to the
external flux:
\begin{equation}\label{CurrA}
    I=\frac{\partial E_n}{\partial \Phi_X}
\end{equation}
This expression can be rewritten in terms of exact Hamiltonian of
a system:
\begin{equation}\label{CurrB}
I = \left\langle n \right|\frac{{\partial \hat H}}{{\partial \Phi
_X }}\left| n \right\rangle
\end{equation}
From (\ref{CurrB}) we would make ansatz  that the current operator
is as follows:
\begin{equation}\label{CurrC}
\Hat{I} = \frac{{\partial \hat H}}{{\partial \Phi _X }}
\end{equation}
However (\ref{CurrC}) is not a consequence of (\ref{CurrB}).
Therefore, the ansatz (\ref{CurrC}) must be proved in every case,
since the current operator in the form of Eq. (\ref{CurrC}) has to
be consistent with its definition in terms of variables of
Hamiltonian $H$. The prove for our case is given below.

The current operator across every junction is a sum of a
supercurrent and a current through the capacitor:
\begin{equation}\label{Curr1}
\hat{I}_i  = I_{Ci} \sin \varphi _{_i }  + \frac{\hbar
}{{2e}}C_i\ddot \varphi _i \quad \left( {i = 1,2} \right)
\end{equation}
We are interested in the current through the inductance coil,
$I_1$ in (\ref{Curr1}) (see Fig.\ref{fig1}).

Direct calculation of $I_1$ with the aid of
(\ref{phidot},\ref{deltadot},\ref{nphidot},\ref{ndeltadot}) yield
the result:
\begin{equation}\label{CurrOp1}
\widehat{I}_1  =  - I_C\frac{{\varphi  - \varphi _X }}{\beta }
\end{equation}
which is independent of parameters of a particular junction in the
loop. From the other hand the expression (\ref{CurrOp1}) can be
obtained from our Hamiltonian (\ref{Ham2}) with the aid of
(\ref{CurrC}). Therefore, the  equation (\ref{CurrC}) gives us the
true expression for the current operator. It is important to note
that the proper expression  for the current operator
(\ref{CurrOp1}) cannot be obtained without magnetic energy term in
the original Lagrangian (\ref{Lag1}).

It follows from (\ref{Ham2}) and (\ref{CurrOp1}) that $\left[
{\hat I,\hat H} \right] \ne 0$. Therefore, an eigenstate of $H$
cannot possess a definite current value.

For the charge qubit the current operator can be obtained from
(\ref{H}) with the aid of its definition (\ref{CurrC}):
\begin{equation}\label{CurrQb}
    \widehat{I}=\frac{2\pi}{\Phi_0}\frac{\partial W}{\partial
    \Phi_X}+\frac{\pi}{\Phi_0}\left[-\frac{B}{\xi}\tau_X+\frac{\xi}{4}A\tau_Y\right]
\end{equation}
The transformation of (\ref{CurrQb}) in the eigenstate basis yeild
the result:
\begin{equation}\label{CurrEig}
    \widehat{S}^{-1}\widehat{I}\widehat{S}=I_0+I_Z\sigma_Z+I_X\sigma_X+I_Y\sigma_Y
\end{equation}
where
\begin{equation}\label{I0}
    I_0=\frac{2\pi}{\Phi_0}\frac{\partial W}{\partial
    \Phi_X}
\end{equation}
\begin{equation}\label{IZ}
I_Z=-\frac{1}{2}\frac{\partial\Delta E}{\partial\Phi_X} =
-\frac{\pi }{{\Phi _0 }}\frac{{E_J^2 }}{{\Delta E}}\left(
{\frac{{\xi ^2 }}{4} - 1} \right)\left( {1 - \frac{{\eta ^2
}}{{16}}} \right)^2 \sin \varphi _X
\end{equation}
\begin{equation}\label{IX}
I_X  = \frac{\pi }{{\Phi _0 }}E_J \left( {1 - \frac{{\eta ^2
}}{{16}}} \right)\sin \frac{{\varphi _X }}{2}\frac{{\left[
{\frac{{\xi ^2 }}{4} + \frac{C}{{\Delta E}}\left( {\frac{{\xi ^2
}}{4} - 1} \right)\cos ^2 \frac{{\varphi _X }}{2}} \right]}}{{\cos
^2 \frac{{\varphi _X }}{2} + \frac{{\xi ^2 }}{4}\sin ^2
\frac{{\varphi _X }}{2}}}
\end{equation}
\begin{equation}\label{IY}
I_Y  =  - \frac{\pi }{{\Phi _0 }}E_J \frac{\xi }{2}\left( {1 -
\frac{{\eta ^2 }}{{16}}} \right)\cos \frac{{\varphi _X
}}{2}\frac{{\left[ {1 + \frac{C}{{\Delta E}}\left( {\frac{{\xi ^2
}}{4} - 1} \right)\sin ^2 \frac{{\varphi _X }}{2}} \right]}}{{\cos
^2 \frac{{\varphi _X }}{2} + \frac{{\xi ^2 }}{4}\sin ^2
\frac{{\varphi _X }}{2}}}
\end{equation}
Therefore, the current operator is not diagonal neither in the
charge basis no in eigenstate basis. If we neglect the inductance
and the current asymmetry ($\beta=0$, $\xi=0$), we obtain:
$I_Y=0$,
\begin{equation}\label{IZ0}
    I_Z =
\frac{\pi }{{\Phi _0 }}\frac{{E_J^2 }}{{\Delta E}} \sin \varphi _X
\end{equation}
\begin{equation}\label{IX0}
    I_X  = -\frac{\pi }{{\Phi _0 }}E_J \frac{C}{{\Delta E}}\sin \frac{{\varphi _X }}{2}
\end{equation}
where
\begin{equation}\label{DE0}
    \Delta E = \sqrt {4E_J^2\cos^2\frac{\varphi _X }{2} +C^2}
\end{equation}
The existence of nondiagonal elements of the current operator in
eigenstate basis is important if we consider the inductive
coupling of several qubits.
\section{Interaction of the charge qubit with its own LC circuit.
Corrections to the qubit energies}

Here we enlarge the Hilbert space to add to two qubit states two
photon states of LC resonator, $a^+a=0, 1$. The transformed
Hamiltonian (\ref{HamTL}), which accounts for transitions between
ground and excited state of LC resonator will read:
\begin{eqnarray}\label{HamQBLC}
 \widehat{S}^{-1}H\widehat{S} = W+Pa^ +  a + Q_1\left( {a^ +   - a} \right)
 + Q_2\left( {a^ +   + a} \right)\\\nonumber-
 \frac{1}{2}\Delta E\sigma _Z
+R\left( {a^ +   - a} \right)\left( {C\sigma _Z  + A\sigma
_X-B\sigma_Y } \right)
 \\\nonumber
  + S\left( {a^ +   + a} \right)\left( {Z\sigma _Z  +X\sigma _X +Y\sigma_Y} \right)
   + Ta^+ a\left( {\sigma _Z-\frac{AC}{\varepsilon^2}\sigma _X  + \frac{BC}{\varepsilon^2}\sigma_Y } \right)
\end{eqnarray}
where
\begin{equation}\label{P}
P = \left[ {E_0  + \frac{{E_J \beta \eta ^2 }}{{32}}\left( {1 +
\frac{{\xi ^2 }}{8} - \left( {1 - \frac{{\xi ^2 }}{8}} \right)\cos
\varphi _X } \right)} \right]
\end{equation}
\begin{equation}\label{Q1}
Q_1  = i\frac{{E_C \alpha }}{{2\sqrt 2 }}\left[ {\gamma  - 2\left(
{1 - \gamma } \right)n_g } \right]
\end{equation}
\begin{equation}\label{Q2}
Q_2  = \frac{{E_J \beta \eta ^3 }}{{128\sqrt 2 }}\left( {1 -
\frac{{\xi ^2 }}{4}} \right)\sin \varphi _X
\end{equation}
\begin{equation}\label{R}
 R =  - \frac{i}{{2\sqrt 2 }}\frac{{E_C \alpha \gamma }}{{\eta \Delta E}}
\end{equation}
\begin{equation}\label{S2}
S = \frac{{\eta ^3 }}{{32\sqrt 2 \xi \left( {1 - \frac{{\eta ^2
}}{{16}}} \right)}}
\end{equation}
\begin{equation}\label{Z}
Z = \frac{{AB}}{{\Delta E}}\left( {\frac{{\xi ^2 }}{4} - 1}
\right)
\end{equation}
\begin{equation}\label{X}
X = B\left( {1 + \frac{{A^2 \left( {\frac{{\xi ^2 }}{4} - 1}
\right)\left( {1 - \frac{C}{{\Delta E}}} \right)}}{{\varepsilon ^2
}}} \right)
\end{equation}
\begin{equation}\label{Y}
Y = A\left( {1 + \frac{{\left( {\frac{{\xi ^2 }}{4} - 1}
\right)\left( {A^2  + B^2 \frac{C}{{\Delta E}}}
\right)}}{{\varepsilon ^2 }}} \right)
\end{equation}
\begin{equation}\label{T}
T = \frac{{\varepsilon ^2 }}{{\Delta E}}\frac{{\eta ^2
}}{{16\left( {1 - \frac{{\eta ^2 }}{{16}}} \right)}}
\end{equation}
The operators $\sigma_X, \sigma_Y, \sigma_Z$ are defined on the
eigenstates $|\uparrow\rangle$, $|\downarrow\rangle$:
\begin{equation}\label{sigma}
\begin{array}{cc}
 \sigma _Z \left|  \uparrow  \right\rangle  = \left|  \uparrow  \right\rangle ;\;\sigma _Z \left|  \downarrow  \right\rangle  =  - \left|  \downarrow  \right\rangle ;\; \\
 \sigma _X \left|  \uparrow  \right\rangle  = \left|  \downarrow  \right\rangle ;\;\sigma _X \left|  \downarrow  \right\rangle  = \left|  \uparrow  \right\rangle ; \\
 \sigma _Y \left|  \uparrow  \right\rangle  = i\left|  \downarrow  \right\rangle ;\;\sigma _Y \left|  \downarrow  \right\rangle  =  - i\left|  \uparrow  \right\rangle\\
 \end{array}
\end{equation}
In addition, we restrict photon subspace to photon numbers n=0,1.
The basis set for our photon+ qubit system is:
\begin{equation}\label{BS}
\begin{array}{cc}
 \left| {0 \uparrow } \right\rangle  = \left| 0 \right\rangle  \otimes \left|  \uparrow  \right\rangle ;\;\;\left| {0 \downarrow } \right\rangle  = \left| 0 \right\rangle  \otimes \left|  \downarrow  \right\rangle ; \\
 \left| {1 \uparrow } \right\rangle  = \left| 1 \right\rangle  \otimes \left|  \uparrow  \right\rangle ;\;\left| {1 \downarrow } \right\rangle  = \left| 1 \right\rangle  \otimes \left|  \downarrow  \right\rangle  \\
 \end{array}
\end{equation}
Within this basis the wave function for Hamiltonian
(\ref{HamQBLC}) is decomposed as:
\begin{equation}\label{WF}
\Psi  = a\left| {0 \uparrow } \right\rangle  + b\left| {0
\downarrow } \right\rangle  + c\left| {1 \uparrow } \right\rangle
+ d\left| {1 \downarrow } \right\rangle
\end{equation}
The Schrodinger equation $H\Psi=E\Psi$ takes the form:
\begin{equation}\label{Schred}
\begin{array}{l}
 \left[ {a\left( {W - \frac{1}{2}\Delta E - E} \right) + c\left( {Q_2  - Q_1  - RC + SZ} \right) - d\left( {RA + iRB - SX + iSY} \right)} \right]\left| {0 \uparrow } \right\rangle  \\
  + \left[ {b\left( {W + \frac{1}{2}\Delta E - E} \right) + c\left( { - RA + iRB + SX + iSY} \right) + d\left( {Q_2  - Q_1  + RC - SZ} \right)} \right]\left| {0 \downarrow } \right\rangle  \\
  + \left[ {a\left( {Q_1  + Q_2  + RC + SZ} \right) + b\left( {RA + iRB + SX - iSY} \right)} \right. \\
 \left. { + c\left( {W + P - \frac{1}{2}\Delta E + T - E} \right) - d\frac{{TC}}{{\varepsilon ^2 }}\left( {A + iB} \right)} \right]\left| {1 \uparrow } \right\rangle  \\
  + \left[ {a\left( {RA - iRB + SX + iSY} \right) + b\left( {Q_1  + Q_2  - RC - SZ} \right)} \right. \\
 \left. { - c\frac{{TC}}{{\varepsilon ^2 }}\left( {A - iB} \right) + d\left( {W + P + \frac{1}{2}\Delta E - T - E} \right)} \right]\left| {1 \downarrow } \right\rangle  = 0 \\
 \end{array}
\end{equation}
The energy levels are defined by equating of the determinant of
this equation to zero. In order to simplify the problem we assume
the inductance of the qubit is very small. In this limit we may
put $W=0$, $P=E_0$, $Q_2=0$, $S=0$, $T=0$. The Hamiltonian
(\ref{HamQBLC}) is reduced to:
\begin{equation}\label{HamRed}
 H = E_0a^ +  a + Q_1\left( {a^ +   - a} \right)-
 \frac{1}{2}\Delta E\sigma _Z
+R\left( {a^ +   - a} \right)\left( {C\sigma _Z  + A\sigma_
X-B\sigma_Y } \right)
\end{equation}

It is important that the inductance cannot be eliminated at all in
the two photon approximation, since the quantity $\beta$ is in the
denominators of $E_0$ and $R$. The Schredinger equation for
Hamiltonian (\ref{HamRed}) is as follows:
\begin{equation}\label{Schred1}
\begin{array}{l}
 \left[ {a\left( { - \frac{1}{2}\Delta E - E} \right) + c\left( { - Q_1  - RC} \right) - dR\left( {A + iB} \right)} \right]\left| {0 \uparrow } \right\rangle  \\
  + \left[ {b\left( {\frac{1}{2}\Delta E - E} \right) + cR\left( { - A + iB} \right) + d\left( { - Q_1  + RC} \right)} \right]\left| {0 \downarrow } \right\rangle  \\
  + \left[ {\left[ {a\left( {Q_1  + RC} \right) + bR\left( {A + iB} \right) + c\left( {E_0  - \frac{1}{2}\Delta E - E} \right)} \right.} \right]\left| {1 \uparrow } \right\rangle  \\
  + \left[ {aR\left( {A - iB} \right) + b\left( {Q_1  - RC} \right) + d\left( {E_0  + \frac{1}{2}\Delta E - E} \right)} \right]\left| {1 \downarrow } \right\rangle  = 0 \\
 \end{array}
\end{equation}
The energy levels are defined by equating of the determinant of
the following matrix to zero:
\begin{equation}\label{Matrix}
\left| {\begin{array}{*{20}c}
   { - \frac{1}{2}\Delta E - E} & 0 & { - Q_1  - RC} & { - R\left( {A + iB} \right)}  \\
   0 & {\frac{1}{2}\Delta E - E} & { - R\left( {A - iB} \right)} & { - Q_1  + RC}  \\
   {Q_1  + RC} & {R\left( {A + iB} \right)} & {E_0  - \frac{1}{2}\Delta E - E} & 0  \\
   {R\left( {A - iB} \right)} & {Q_1  - RC} & 0 & {E_0  + \frac{1}{2}\Delta E - E}  \\
\end{array}} \right| = 0
\end{equation}

Below we calculate the energies for the following set of the qubit
parameters: $E_J=4.64\times10^{-24}J$, $E_C=10E_J$, $n_g=-0.5$;
$\xi=0.1$. The calculations have been performed for two cases. For
noninteracting qubit we used the expression (\ref{DE}), where the
LC circuit only renormalizes the energy due to the vacuum
fluctuations of LC ocsillator (factor $\eta$ in (\ref{DE})). For
the qubit which interacts with its own LC circuit we solved the
equation (\ref{Matrix}).

\begin{figure}[tbp]
  \centerline{\includegraphics[width=7.5cm, angle=-90]{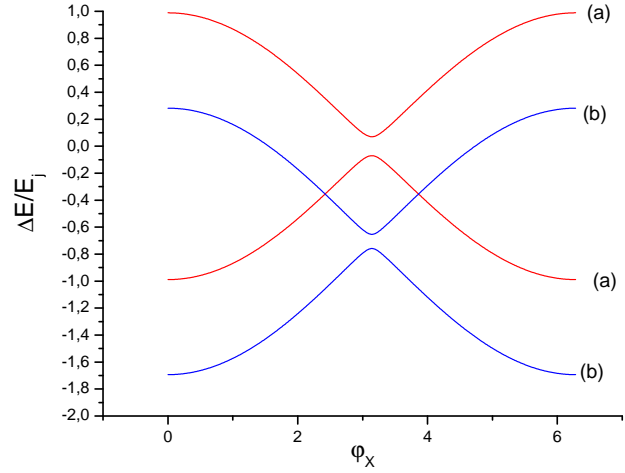}}
  \caption{The plots of qubit energies vs magnetic flux $\phi_X$ for $\gamma=0.01$,
  $\beta=0.001$. The red curves (a) are the ground and excited states for a qubit
  which does not interact with its LC resonator. The blue curves (b) are the same as (a)
  but for a qubit which interacts with its LC resonator.}\label{fig2}
\end{figure}
\begin{figure}[tbp]
  \includegraphics[width=7.5cm, angle=-90]{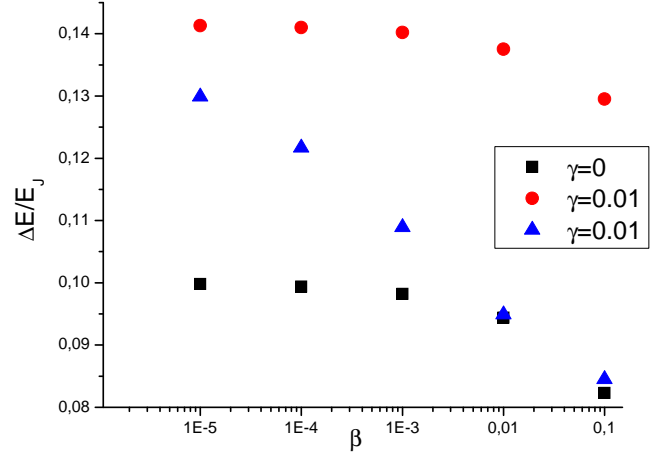}\\
  \caption{The dependance of minimum gap on $\beta$. The black boxes and red circles are
  calculated for noninteracting qubit. The blue triangles are those for interacting qubit.}\label{fig3}
\end{figure}
The plots of the qubit energy levels are shown on Fig.\ref{fig2}.
As is seen from these plots the finite value of $\beta$, though it
is rather small, modifies the gap between ground and first excited
energy levels. The calculations show that at the point
$\phi_X=\pi$, where the gap is minimum, the gap for noninteracting
qubit is $\Delta E=0.14E_J$. The interaction of the qubit with its
LC circuit reduces the gap to $\Delta E=0.11E_J$.

The dependance of minimum gap (at the point $n_g=-0.5$,
$\phi_X=\pi$) on $\beta$ is shown on Fig.\ref{fig3}. As is seen
from Fig.\ref{fig3} the noninteracting qubit is slightly modified
by the inductance. A small reduction of the gap with the increase
of $\beta$ is due to the factor $\eta$ in (\ref{DE}). However, the
reduction of the gap with the increase of $\beta$ for interacting
qubit is much more significant (black triangles on
Fig.\ref{fig3}). It is important to note that this effect is more
pronounced for relative large $\gamma$'s. For $\gamma=0$ the
interaction with LC circuit does not alter the energies (compared
to those for noninteracting case).

\textbf{Acknowledgements}

Ya. S. Greenberg acknowledges the financial support from Deutsche
Forschungsgemeinschaft.

\end{document}